\documentclass[a4paper,aps,prl,showpacs,twocolumn]{revtex4-1}

\usepackage[utf8]{inputenc}
\usepackage[T1]{fontenc}
\usepackage[british]{babel}
\usepackage[scaled=1.03]{inconsolata}
\usepackage{xcolor,lmodern,tikz,physics,graphicx,amsmath,amssymb,amsthm,bm,mathtools,amsfonts,mathrsfs,bbm,dsfont}
\definecolor{darkred}{rgb}{0.5,0,0}
\definecolor{darkgreen}{rgb}{0,0.5,0}
\definecolor{darkblue}{rgb}{0,0,0.5}
\usepackage[colorlinks,breaklinks,linkcolor=darkred,citecolor=darkgreen,urlcolor=darkblue]{hyperref}
\usepackage[caption=false]{subfig}
\usepackage[babel]{microtype}
\newcommand{\id}{\ensuremath{\mathds{1}}}
\newcommand{\meas}[1]{\ensuremath{\mathcal{#1}}}

\newcommand{\rk}{\ensuremath{\text{rk}}}
\renewcommand{\leq}{\leqslant}
\renewcommand{\geq}{\geqslant}

\newtheoremstyle{mystyle}{6pt}{6pt}{\normalfont}{0pt}{\bf}{.}{ }{}
\theoremstyle{mystyle}
\newtheorem{observation}{Observation}

\graphicspath{{figs/}}

\begin{document}
\nonfrenchspacing

\title{Quantum Entanglement in the Triangle Network}

\author{Tristan Kraft,$^1$ S\'ebastien Designolle,$^2$ Christina Ritz,$^1$ Nicolas Brunner,$^2$ Otfried G\"uhne,$^1$ Marcus Huber$^3$}
\affiliation{
  \makebox[0pt]{$^1$Naturwissenschaftlich-Technische Fakultät, Universität Siegen, Walter-Flex-Straße 3, 57068 Siegen, Germany}\\
  \makebox[0pt]{$^2$Department of Applied Physics, University of Geneva, 1211 Geneva, Switzerland}\\
  \makebox[0pt]{$^3$Institute for Quantum Optics and Quantum Information, Austrian Academy of Sciences, 1090 Vienna, Austria}
}

\date{\today}

\begin{abstract}
  Beyond future applications, quantum networks open interesting fundamental perspectives, notably novel forms of quantum correlations.
  In this work we discuss quantum correlations in networks from the perspective of the underlying quantum states and their entanglement.
  We address the questions of which states can be prepared in the so-called triangle network, consisting of three nodes connected pairwise by three sources.
  We derive necessary criteria for a state to be preparable in such a network, considering both the cases where the sources are statistically independent and classically correlated.
  This shows that the network structure imposes strong and non-trivial constraints on the set of preparable states, fundamentally different from the standard characterization of multipartite quantum entanglement.
\end{abstract}

\maketitle

{\it Introduction.---}
Advances in quantum information processing and technologies lead to promising developments towards a quantum network, see, e.g.,~\cite{Kimble2008,Sangouard2011,Simon2017,Wehner2018}.
The latter would feature local quantum processors exchanging information and entanglement via quantum links, enabling, for instance, long-distance quantum communication.
While this represents an outstanding technological challenge, recent works have already reported the implementation of basic quantum networks nodes, based on physical platforms where light and matter interact~\cite{Cirac1997,Duan2001,Tanzilli2005,Chaneli2005}.

These developments also raise important questions on the theoretical level.
In the spirit of quantum teleportation or entanglement swapping, the entanglement initially generated on the links of the networks can then be propagated to the entire network by performing entangled measurements at the nodes, see, e.g.,~\cite{Acin2007,Perseguers2008a,Perseguers2008b}.
This effect may lead to extremely strong forms of multipartite quantum correlations, spread across the whole network.
Characterizing such correlations is a natural question, of clear fundamental interest, but which may also impact the development of future experimental quantum networks.

\begin{figure*}
  \subfloat[\label{sfig:delta_I}]{
    \includegraphics[width=.15\linewidth]{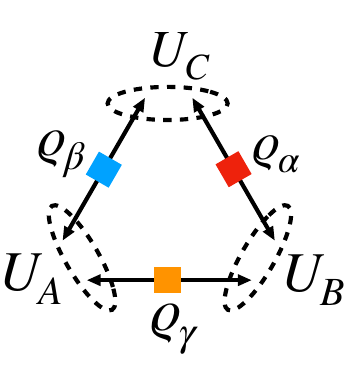}
  }
  \hfill
  \subfloat[\label{sfig:delta_C}]{
    \includegraphics[width=.18\linewidth]{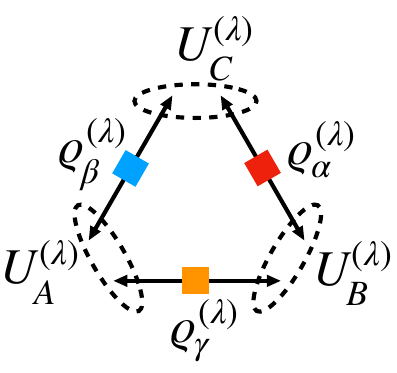}
  }
  \hfill
  \subfloat[\label{sfig:sep}]{
    \includegraphics[width=.13\linewidth]{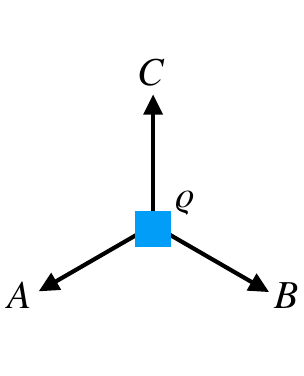}
  }
  \hfill
  \subfloat[\label{sfig:statespace}]{
    \includegraphics[width=.32\linewidth]{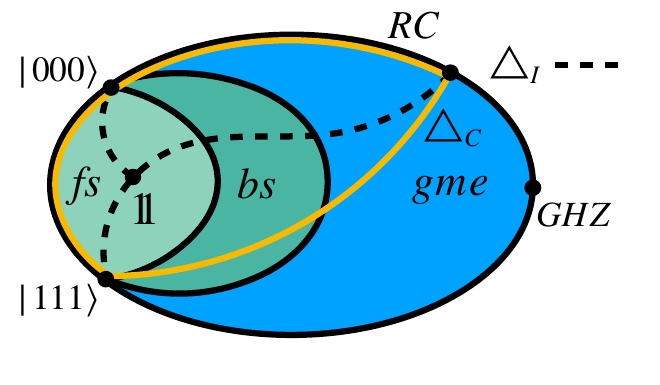}
  }
  \caption{This work discusses the generation of tripartite quantum states in the triangle network.
    We first consider the independent triangle network, shown in (a), where all three quantum sources (producing the bipartite states $\varrho_\alpha$, $\varrho_\beta$, and $\varrho_\gamma$) are statistically independent.
    Each party, upon receiving two independent subsystems, can perform a local unitary.
    We also consider the correlated triangle network, shown in (b), where all sources and nodes are classically correlated via a shared random variable $\lambda$.
    We derive a number of criteria for characterizing which tripartite quantum states $\rho$ can be prepared in each of these scenarios.
    Notably, this problem is fundamentally different from the standard classification of multipartite quantum states, where all nodes share a quantum state distributed from a single common source, as in (c). Comparing the three scenarios one arrives at (d). The green area contains fully separable (fs) states and biseparable (bs) states. The blue area highlights genuine multipartite entangled (gme) states. The dashed line corresponds to states contained in $\Delta_I$, e.g., the ring cluster state (RC) defined in the main text. The orange line corresponds to the boundary of $\Delta_C$}.
  \label{fig:network}
\end{figure*}

First steps have been taken towards characterizing quantum nonlocality in networks.
To do so, the concept of Bell locality~\cite{Bell1964} has been generalized to networks~\cite{Branciard2010,Rosset2016,Fritz2012}.
The key idea is that the different sources in the network distributing physical systems to the nodes should be assumed to be independent from each other.
This represents a fundamental departure from standard Bell nonlocality, and new striking effects can occur.
For instance, it is possible to detect quantum nonlocality in an experiment involving fixed measurements, i.e., a Bell inequality violation ``without inputs''~\cite{Fritz2012,Branciard2012,Fraser2018}, as well as novel forms of quantum correlations genuine to networks~\cite{Renou2019}.
While significant progress has been reported in recent years, see, e.g.,~\cite{Henson2014,Chaves2015,Wolfe2019,Renou2019b,Pozas2019}, as well as first experiments~\cite{Saunders2017,Carvacho2017,Sun2019}, the study of these phenomena is still in its infancy, and it is fair to say that our understanding of quantum correlations in networks remains very limited so far.

The focus of the present work is to investigate quantum correlations in networks from the point of view of entanglement.
Inspired from the developments above and recent developments in entanglement theory~\cite{Kraft2018}, we discuss the generation of multipartite entangled states in a network.
We focus our attention on the so-called ``triangle network'', which is known to exhibit interesting forms of quantum nonlocal correlations~\cite{Fritz2012,Renou2019}.
This simple network features three nodes, each pair of nodes being connected by a bipartite quantum source (see Fig.~\ref{fig:network}).
We explore the possibilities and limits for entanglement generation, given the constraints of the network topology.
Notably, it turns out that not all quantum states can be prepared.
We discuss two scenarios, featuring independent or classically correlated quantum sources and unitaries, and derive general conditions for a quantum state to be preparable in the network.
This allows us to show that important classes of multipartite quantum states cannot be prepared in the triangle network, including also some separable states in the case of independent sources.
On the other hand, certain genuinely multipartite entangled states can be created in the network.
This shows that the network structure imposes strong and nontrivial constraints on the set of possible quantum states.
Our work represents a first step towards understanding quantum correlations in networks from the point of view of quantum states and their entanglement.

{\it Triangle network with independent sources.---}
We consider a simple network featuring three nodes: $A$, $B$ and $C$.
These nodes are connected pairwise by three sources, hence the network forms a triangle.
Each source produces a bipartite quantum state (of arbitrary dimension $d \times d$): $\varrho_\alpha$ is shared by $B$ and $C$, $\varrho_\beta$ by $A$ and $C$, and $\varrho_\gamma$ by $A$ and $B$.
Thus, each party receives two $d$-dimensional quantum systems.
Finally, each party can apply a local unitary to their two-qudit systems, which we denote with $U_A$, $U_B$, and $U_C$.
This results in a global state $\varrho$ for the network, see Fig.~\ref{sfig:delta_I}.
Note that any state of the triangle network can be mapped to a three qudit state with local dimension $d^2$.
E.g., for $d=2$ we use $\ket{00}\rightarrow\ket{0}$, $\ket{01}\rightarrow\ket{1}$, $\ket{10}\rightarrow\ket{2}$ and $\ket{11}\rightarrow\ket{3}$, and we will refer to this as the standard encoding.

In the first part of this paper, we will focus on the scenario where the three sources are assumed to be statistically \emph{independent} from each other.
This we call the independent triangle network (ITN), and the set of states that can be prepared by such a network we denote by $\bigtriangleup_I$.
Statistical independence of the sources is a relatively natural starting point for building an entanglement theory for practical quantum networks, analogous to product states being the free resource of separate parties with only local sources of quantum states. Additionally, one may then generalise to include classically correlated parties (sources and nodes), which is a scenario that we discuss later in the paper.

The first question we consider is which quantum states $\rho$ can be prepared in the ITN.
Specifically, we say that $\rho \in \bigtriangleup_I$ if it admits a decomposition of the form: 
\begin{equation}\label{eqn:triangle_ind}
  \rho = (U_A\otimes U_B\otimes U_C)(\varrho_\alpha\otimes\varrho_\beta\otimes\varrho_\gamma)(U_A^\dagger\otimes U_B^\dagger\otimes U_C^\dagger).
\end{equation}
Note that here we use a compact notation where the order to the sub-systems is not the same in the unitaries and in the states.

As intuition suggests, there exist tripartite quantum states that cannot be prepared in the triangle network.
In fact, $\bigtriangleup_I$ represents only a zero-measure subset of the entire set of quantum states in $\mathcal{H}$, as confirmed by counting the free parameters \footnote{For $\rho \in \bigtriangleup_I$, we have $6(d^2-1)$ parameters (count $d^2-1$ per state and per unitary), which is indeed is much smaller than the $d^{12}-1$ parameters for a general state in $\mathcal{H}=\mathcal{L}(\mathbb{C}^{d^6})$.}.
In the following we discuss the characterization of $\bigtriangleup_I$ which is challenging, mainly due to the fact it is a nonconvex set, as we will see below.

{\it Preparability in the ITN.---}
We now present three different criteria that give necessary conditions satisfied by any $\rho \in \bigtriangleup_I$.
They capture limits on classical and quantum correlations for such states, as well as restriction on ranks.
We first present the criteria, and then apply them to illustrative examples.

From Fig.~\ref{sfig:delta_I} it appears clear that the amount of global classical correlations for any $\rho \in \bigtriangleup_I$ must be limited.
Indeed, the three nodes do not share any common (i.e., tripartite) information.
This intuition can be made formal by considering the so-called tripartite mutual information for quantum systems (TMI)~\cite{Cerf1998}.
It is defined as $I_{3}(A:B:C)=I_{2}(A:B)+I_{2}(A:C)-I_{2}(A:BC)$, where $I_{2}(X:Y)=S(X)-S(X\vert Y)=S(X)+S(Y)-S(X,Y)$ is the bipartite quantum mutual information, and $S(\cdot)$, $S(\cdot\vert\cdot)$, and $S(\cdot,\cdot)$ are the von Neumann usual, conditional, and joint entropies, respectively.
Then the TMI reads
\begin{equation}\label{eqn:tmi}
  \begin{aligned}
    I_3(A:B:C)=&\,S(ABC)+S(A)+S(B)+S(C)\\&-S(AB)-S(AC)-S(BC).
  \end{aligned}
\end{equation}
Since the von Neumann entropy is invariant under unitary transformations and additive on tensor products, it follows form Eq.~\eqref{eqn:triangle_ind} that  $S(\rho)=S(\varrho_\alpha)+S(\varrho_\beta)+S(\varrho_\gamma)$ for any $\rho \in \bigtriangleup_I$.
Expanding the bipartite entropies as, e.g., $S(AB)=S(\tr_C\varrho_\beta)+S(\tr_C\varrho_\alpha)+S(\varrho_\gamma)$, we arrive at the following:
\begin{observation}\label{obs:tmi}
  $I_3(A:B:C)=0$ for any $\rho \in \bigtriangleup_I$.
\end{observation}

Moving beyond classical correlations, we now observe that quantum correlations are also limited for states in $\bigtriangleup_I$.
From Fig.~\ref{sfig:delta_I}, the intuition is that the entanglement on the bipartition $A\vert BC$ should be equal to the sum of the entanglement in the reduced states, i.e., $A\vert B$ and $A\vert C$.

This can be shown formally by using an appropriate entanglement measure.
Recall that a quantity $\meas{E}[\sigma]$ is called an entanglement measure, if (i) $\meas{E}[\sigma]$ vanishes for separable states, (ii) it does not change under local choice of basis, i.e., local unitary transformations, and (iii) it does not increase (on average) under local operations and classical communication.
In our case, we further require two properties.
First, the measure must be additive on tensor products, i.e., $\meas{E}[\sigma_1\otimes\sigma_2]=\meas{E}[\sigma_1]+\meas{E}[\sigma_2]$.
Second, we need the so-called monogamy constraint:
\begin{equation}\label{eqn:monogamy}
  \meas{E}_{X\vert Y}[\sigma_{XY}]+\meas{E}_{X\vert Z}[\sigma_{XZ}]\leq\meas{E}_{X\vert YZ}[\sigma_{XYZ}],
\end{equation}
where, e.g., $\sigma_{XY}=\tr_{Z}\sigma_{XYZ}$ denotes a reduced state.
An example of such an entanglement measure is the squashed entanglement~\cite{Christandl2004,Koashi2004}; note however that not all entanglement measures satisfy the above properties, see, e.g., Refs~\cite{Plenio2007,Guehne2009}.

For states in $\bigtriangleup_I$, we first note that the local unitaries $U_A$, $U_B$ and $U_C$ can always be disregarded, since they do not change the amount of entanglement between the parties.
Hence the right-hand side of Eq.~\eqref{eqn:monogamy} can be evaluated as $\meas{E}_{A\vert BC}=\meas{E}_{A_\beta A_\gamma\vert B_\alpha B_\gamma C_\alpha C_\beta}=\meas{E}_{A_\gamma\vert B_\gamma}+\meas{E}_{A_\beta\vert C_\beta}= \meas{E}_{A\vert B}+\meas{E}_{A\vert C}$, where $A_\beta$ denotes the subsystem that $A$ receives from the source $\beta$ (connecting nodes $A$ and $C$), and similarly for other subsystems.
Thus, we arrive at the following:
\begin{observation}\label{obs:ent}
  Let $\meas{E}[\cdot]$ be an entanglement measure that is additive on tensor products and monogamous.
  For any $\rho \in \bigtriangleup_I$ we have that $\meas{E}_{X\vert YZ}\qty[\rho]= \meas{E}_{X\vert Y}\qty[\tr_Z\rho]+\meas{E}_{X\vert Z}\qty[\tr_Y\rho]$ holds for all the bipartitions $A\vert BC$, $B\vert AC$ and $C\vert AB$.
\end{observation}

Finally, we show that the structure of the ITN imposes constraints on the ranks of the global state and of its marginals.
For $\rho \in \bigtriangleup_I$, we have from Eq.~\eqref{eqn:triangle_ind} that $\rk(\rho)=\rk(\varrho_\alpha)\rk(\varrho_\beta)\rk(\varrho_\gamma)$ since the unitaries $U_A$, $U_B$, and $U_C$ do not affect the global rank.
Likewise, the ranks of the local reduced states satisfy $\rk(\tr_{BC}\rho)=\rk(\tr_C\varrho_\beta)\rk(\tr_B\varrho_\gamma)$, and those of the bipartite reduced states $\rk(\tr_C\rho)=\rk(\tr_C\varrho_\beta)\rk(\varrho_\gamma)\rk(\tr_C\varrho_\alpha)$, and similarly for the other marginals.
Therefore we get the following observation:
\begin{observation}\label{obs:rank}
  For $\rho \in \bigtriangleup_I$ there exist integers $r_\alpha$, $r_\beta$, $r_\gamma$ in $[1,d^2]$ and $r_\gamma^A$, $r_\gamma^B$, $r_\alpha^B$, $r_\alpha^C$, $r_\beta^C$, $r_\beta^A$ in $[1,d]$ such that
  \vspace{-0.5cm}
  \begin{equation}
    \begin{array}{c}
      \rk(\rho)=r_\alpha r_\beta r_\gamma,\\[4pt]
      \rk(\tr_A\rho)=r_\alpha r_\beta^Cr_\gamma^B,\quad
      \rk(\tr_{BC}\rho)=r_\beta^Ar_\gamma^A,\\[4pt]
      \rk(\tr_B\rho)=r_\alpha^Cr_\beta r_\gamma^A,\quad
      \rk(\tr_{AC}\rho)=r_\alpha^Br_\gamma^B,\\[4pt]
      \rk(\tr_C\rho)=r_\alpha^Br_\beta^Ar_\gamma,\quad
      \rk(\tr_{AB}\rho)=r_\alpha^Cr_\beta^C.
    \end{array}
  \end{equation}
\end{observation}
Furthermore, it is worth noting that in the case of pure states $\ket{\psi}$ a prime tensor rank of the state does not exclude the possibility of the state $\ket{\psi}$ being preparable in the ITN (see Appendix, Sec.~I for more details).

{\it Illustrative examples.---}
To demonstrate the relevance of the above criteria, we now discuss some examples, which highlight some interesting properties of the set $\bigtriangleup_I$.

Consider first the classically correlated state defined by $\rho_{k}^{C} = \frac{1}{k}\sum_{j=0}^{k-1} \ket{jjj}\bra{jjj}$, with $1 \leq k \leq d^2$.
For $k=1$, we have simply a product state, hence the state is trivially preparable in the ITN.
However, for any $k\geq2$, it follows from Observation~\ref{obs:tmi} that $\rho_{k}^{C}$ is outside of $\bigtriangleup_I$, as ${I_3(A:B:C)=\log_2(k)\neq 0}$.
This shows that the set of states $\bigtriangleup_I$ is nonconvex (see Fig.~\ref{sfig:statespace}).
Moreover, note that $\rho_{k}^{C}$ is fully separable (in the usual approach to multipartite entanglement, see Fig.~\ref{sfig:sep}).

Furthermore, consider the tripartite GHZ state of local dimension $d^2$: $\ket{GHZ_{d^2}} = \frac{1}{d}\sum_{j=0}^{d^2-1} \ket{jjj}$.
From Observation~\ref{obs:ent}, we see that the state does not belong to $\bigtriangleup_I$.
More dramatically, consider a noisy GHZ state of the form $ \rho_V = V \ket{GHZ}\bra{GHZ} + (1-V)\id/d^6$, where $V \in [0,1]$ is the visibility.
From Observation~\ref{obs:tmi}, it follows that $\rho_V \in \bigtriangleup_I$ only if $V=0$.
Hence, there exist states arbitrarily close to the fully mixed state, that yet are outside $\bigtriangleup_I$.

These examples illustrate that the characterization of $\bigtriangleup_I$ is completely different from the standard characterization of tripartite quantum state, distributed by a single common source (as in Fig.~\ref{sfig:sep}).
First, there exist fully separable tripartite states, such as $\rho_{k}^{C}$, that are not in $\bigtriangleup_I$.
Moreover, such states can be found arbitrarily close to the fully mixed state (see Fig.~\ref{sfig:statespace}).

However, there also exist highly entangled states within $\bigtriangleup_I$.
An example is the six-qubit ring cluster (RC) state~\cite{Hein2004}.This state can be prepared in the ITN: each source generates a maximally entangled two-qubit state, and each party applies a controlled-$\sigma_z$ unitary.
This state is nonetheless entangled in the strongest sense, as it features genuine tripartite entanglement~\cite{Guehne2009}.

Finally, it is natural to ask whether one (or more) of the above three observations could be tight, i.e., a necessary and sufficient condition for membership in $\bigtriangleup_I$.
Clearly, Observation~\ref{obs:ent} cannot be tight: this criterion is based on the entanglement on bipartitions, but there exist fully separable states outside $\bigtriangleup_I$.
Observation~\ref{obs:rank} cannot be tight either: indeed, this criterion can only rule out non-full-rank states.
Lastly, while Observation~\ref{obs:tmi} is also not tight, since $I_3(A:B:C)=0$ for all tripartite pure states, one can obtain a stronger condition by allowing for local quantum channels to be performed at each node.
\begin{observation}
For any pair of local channels one finds that the TMI cannot be rendered positive for any state in $\Delta_I$, e.g.,
$I_3(\Lambda_A\otimes\Lambda_B\otimes\id (\varrho))\leq 0$ for all channels $\Lambda_A,\Lambda_B$ and $\varrho\in\Delta_I$.
\end{observation}
One can easily verify that this condition reduces to the monotonicity property of the bipartite mutual information (see Appendix~Sec.~IV). Using for instance local depolarizing maps would then show that the pure GHZ state is outside $\bigtriangleup_I$, since the depolarized GHZ state has $I_3(A:B:C)>0$. For three channels, such a condition is no longer true, since the TMI can be rendered positive even in the case of classical triangle networks~\cite{Weilenmann2018}.

{\it Triangle network with classical correlations.---}
Moving on from uncorrelated sources, we now consider the scenario where the three nodes and the three sources are classically correlated, e.g., via a common central source of shared randomness (see Fig.~\ref{sfig:delta_C}). This is analogous to separable mixed states being a convex combination of product states, i.e. one way of defining them would be through classically correlated sources of product states.
This we call the correlated triangle network (CTN), and the set of states that can be prepared by such a network we denote by $\bigtriangleup_C$.
A state $\rho$ is feasible in $\bigtriangleup_C$ if it admits a decomposition of the form 
\begin{equation}\label{eqn:triangle_cvx}
  \rho = \sum_\lambda p_\lambda  \rho_\lambda,
\end{equation}
where $\rho_\lambda \in \bigtriangleup_I$, i.e., each $\rho_\lambda$ admits a decomposition of the form~\eqref{eqn:triangle_ind}, and $\lambda$ represents the classical variable shared by all parties (nodes and sources), with density $p_\lambda$.
Hence the set of states $\bigtriangleup_C$ is simply given by the convex hull of the set $\bigtriangleup_I$ (see Fig.~\ref{sfig:statespace}).
While any state in $\bigtriangleup_I$ is also trivially in $\bigtriangleup_C$, the converse is not true as $\bigtriangleup_I$ is nonconvex, as we saw before.
Let us now discuss the properties of $\bigtriangleup_C$.
First, we observe that not all states can be prepared in the CTN.
In particular we make the following observation.
\begin{observation}\label{obs:222gme}
  No three-qubit genuine multipartite entangled state, embedded in larger dimensional systems, can be prepared in the CTN.
\end{observation}
First, we prove the statement for pure states.
To do so, note that the rank of the global state is one and it is entangled along each bipartition.
Hence, due to the Schmidt decomposition, all single party reduced states have rank two.
This, however, is impossible in the ITN.
Recall, that the local ranks are determined by the sources only.
Thus, if one source prepares a two-qubit entangled state the local ranks at the connected nodes are two and the remaining one has rank one.
If two sources produce a two-qubit entangled state there is one reduced state which has rank four, which proves the claim.
Furthermore, note that also no mixed three-qubit genuine multipartite entangled state can be prepared in the CTN.
Such a mixed state necessarily has a pure three-qubit genuine multipartite entangled state in its range and is thus not preparable in the CTN.

Observation~\ref{obs:222gme} in itself is already quite interesting, since it rules out a large class of states that are not preparable in the CTN.
However, the set $\bigtriangleup_C$ can be characterized in a more refined way.
To do that we can take advantage of the convexity of $\bigtriangleup_C$ in order to characterize the set efficiently using numerical methods.
Borrowing techniques from entanglement witnesses~\cite{Guehne2009} and Ref.~\cite{Kraft2018} we now construct ``preparability witnesses'' for determining whether a state belongs to $\bigtriangleup_C$ or not.

Consider a target pure state $\ket{\psi}$ (typically not in $\bigtriangleup_C$).
Define the linear operator $W=\mu^2 \mathds{1}-\ketbra{\psi}$, where $\mu$ denotes the largest  overlap between $\ket{\psi}$ and any state in $\bigtriangleup_I$.
The challenge is now to estimate $\mu$, i.e., to find the maximal overlap between $\ket{\psi}$ and any $\rho \in \bigtriangleup_C$.
For this, it is sufficient to consider pure states in Eq.~\eqref{eqn:triangle_ind}, namely, $\ket{\varphi}=(U_{A}\otimes U_{B} \otimes U_{C} )\ket{\alpha}\otimes\ket{\beta}\otimes\ket{\gamma}$.
In order to compute
\begin{equation}\label{eqn:fidelity}
  \mu = \max_{\substack{U_A,U_B,U_C \\ \ket{\alpha}, \ket{\beta}, \ket{\gamma}}} \abs{\bra{\alpha\beta\gamma}\big(U_A\otimes U_B \otimes U_C\big)\ket{\psi}},
\end{equation}
we perform a see-saw numerical optimization procedure.
We start with random states $\ket{\alpha}$, $\ket{\beta}$, and $\ket{\gamma}$ and random unitaries $U_{A}$, $U_{B}$, and $U_{C}$, where the dimension $d$ of the sources is chosen large enough so that the state $\ket{\psi}$ can be embedded into the space of local dimension $d^2$.
Then we optimize over each state and each unitary one by one, while keeping everything else fixed (see Appendix Sec.~II for details of the algorithm as well as analytical upper bounds on the overlap).
Although we are not guaranteed to find the global maximum, we found that in practice the method works well for low dimensions.
In Table~\ref{tab:fidelities} we give results for some states of interest.

For instance, we can consider a GHZ state of local dimension $d^2=4$ as target state and use the standard encoding to embed this state into the triangle network.
For this we numerically found $\mu^2=1/2$.
This is obtained by choosing all three states to be two-qubit Bell states,  $\ket{\alpha}=\ket{\beta}=\ket{\gamma}=\frac{1}{\sqrt{2}}\qty(\ket{00}+\ket{11})$, and local unitaries $U_A=\text{SWAP}_{A_\beta A_\gamma}$, $U_B=\text{CNOT}_{B_\gamma B_\alpha}$, and $U_C=\text{CNOT}_{C_\beta C_\alpha}$, where the first qubit is the control qubit.

\begin{table}
  \centering
  \begin{tabular}{c|cccccccccccc}
    State $\ket{\psi}$ && $GHZ_2$ && $GHZ_3$ && $GHZ_4$ && $W$ && $AME$ && $AS_3$\\  \hline 
    Putative $\mu^2$ && $\frac{1}{2}$ && $\frac{4}{9}$ && $\frac12$ && $\frac{6}{9}$ && $\frac{1}{2}$ && $0.5362(5)$
  \end{tabular}
  \caption{
    Results of the see-saw algorithm to compute a lower bound on $\mu^2$ given in Eq.~\eqref{eqn:fidelity} for different target states $\ket{\psi}$.
    $AME$ is the absolutely maximally entangled state of six qubits (three ququarts) and $AS_3$ is the totally antisymmetric state on three qutrits~\cite{Cabello2002,Jex2003}.
    All states, except the $AME$ state, are embedded into the triangle network by choosing local dimension $d^2=4$ and using the standard encoding.
  }
  \label{tab:fidelities}
\end{table}

Therefore we see that the set $\bigtriangleup_C$, while being now of full measure and containing all fully separable tripartite states, is still a strict subset of the set of all tripartite quantum states.
It would be interesting to understand which quantum state is furthest away from $\bigtriangleup_C$; our numerical results suggest that these could be of the GHZ form.

Finally, note that one could also consider the scenario where only the three sources are connected by the shared variable $\lambda$, and not the unitaries.
We give preliminary results for this scenario in the Appendix Sec.~III.

{\it Conclusions.---}
In this manuscript we have discussed the structure of quantum states in a simple triangle network.
We identified a number of properties of such states, notably limits on their classical and quantum correlations.
This allowed us to derive necessary criteria for a state to be feasible in the network, as well as witnesses for detecting states that are not preparable. It would be interesting to derive stronger criteria, perhaps even necessary and sufficient, for states in the ITN and/or in the CTN. Another direction is to quantify entanglement in such a network.

More generally, our work represents a natural starting point for building a general theory of network entanglement. Of particular interest are networks involving many nodes connected by bipartite sources, which correspond to current experimental prospects of quantum networks. We note that all the methods presented here can be directly generalized to such class of networks. This could provide a preliminary, yet still valuable information about those yet unexplored structures. Finally, it would be interesting to explore the connection between our approach and quantum causal modelling \cite{Pienaar2020,Barrett2019}.

\begin{acknowledgements}
  We thank Joe Bowles, Flavien Hirsch, Ivan \v{S}upi\'{c}, Goh Koon Tong, Cornelia Spee, and Zhen-Peng Xu for discussions.
  This work was supported by the DFG, the ERC (Consolidator Grant No.~683107/TempoQ) and the Swiss National Science Foundation (Starting Grant DIAQ and NCCR-QSIT).
  MH acknowledges funding from the Austrian Science Fund (FWF) through the START project Y879-N27.

  {\it Note added.---}
  While finishing this manuscript, we became aware of a related works by Navascues et al.~\cite{Navascues2020} and Luo~\cite{Luo2020}.
\end{acknowledgements}

\bibliography{KDR+20}
\bibliographystyle{sd2}

%\appendix
\section{I. Prime tensor rank (Schmidt measure) of a pure state does not imply nonpreparability in the triangle network}
\label{app:prime}

For multipartite pure states their degree of entanglement can be characterized by the so-called Schmidt measure that was introduced in Ref.~\cite{Eisert2001}, which is equivalent to the tensor rank of the coefficient tensor of a pure state $\ket{\psi}$.
Namely, it is the smallest number $r$ of product terms such that $\ket{\psi}=\sum_{i=1}^r \alpha_i\ket*{\psi_{A_1}^{(i)}}\otimes\ket*{\psi_{A_2}^{(i)}}\otimes\cdots\otimes\ket*{\psi_{A_n}^{(i)}}$, where $n$ is the number of parties.
In the bipartite case this reduces to the Schmidt rank~\cite{Guehne2009}.

In the scenario that was considered in Ref.~\cite{Kraft2018} it was proven that if a pure state has prime tensor rank, it cannot be decomposed into lower-dimensional states, e.g., the GHZ state on three ququarts can be decomposed into two two-qubit GHZ states, whereas the GHZ on three qutrits cannot be decomposed into lower-dimensional systems since its tensor rank is three.
In the triangle network this is no longer true, namely, a prime tensor rank of a pure state does not imply that it cannot be produced in the ITN.
Consider a network state, where each source prepares a two-qubit maximally entangled state $\ket{\psi^+}=\frac{1}{\sqrt{2}}\qty(\ket{00}+\ket{11})$.
This state corresponds to the tensor
\begin{equation}
  T=\sum_{i,j,k=0}^1(\ket{i}\otimes\ket{j})\otimes(\ket{j}\otimes\ket{k})\otimes(\ket{k}\otimes\ket{i}),
\end{equation}
which is known as the two-by-two matrix multiplication tensor~\cite{Christandl2019}.
In this decomposition the tensor can be represented as a sum of $2^3=8$ terms.
However, it is known that this tensor has tensor rank seven~\cite{Strassen1969}.
From that we conclude that, although the state has a prime tensor rank, it can be prepared in the triangle network, and hence, a prime tensor rank does not mean that a state is not preparable in the triangle network.

\section{II. Algorithm and analytical estimates for the maximal overlap of a given state with the ITN}
\label{app:algo}

In order to perform the optimization over the states $\ket{\alpha}$, $\ket{\beta}$, and $\ket{\gamma}$, we fix the unitaries and two of the states, say $\ket{\beta}$ and $\ket{\gamma}$.
The state $\ket{\alpha}$ that maximizes the expression $\max_{\ket{\alpha}} \abs*{\bra{\alpha} [\bra{\beta\gamma}\ket*{\widetilde{\psi}}]}$ is simply given by $\ket{\alpha}^{\mathrm{opt}} = \bra*{\beta\gamma}\ket*{\widetilde{\psi}}/N$, where $N$ is a normalization factor and $\ket*{\widetilde{\psi}}$ is the target state including the local unitaries.

In order to perform the optimization over the unitaries $U_A$, $U_B$, and $U_C$, we fix the states and two of the unitaries, say $U_B$ and $U_C$.
To get the optimal choice for $U_A$ we compute $\max_{U_A}\abs*{\bra{\alpha\beta\gamma}U_A\ket*{\widetilde{\psi}}}$, where $\ket*{\widetilde{\psi}}=U_B\otimes U_C\ket{\psi}$.
We can rewrite this as
\begin{equation}
  \max_{U_A}\left|\tr(U_A\ketbra*{\widetilde{\psi}}{\alpha\beta\gamma})\right|=\max_{U_A}\abs{\tr_A(U_A\rho_A)},
\end{equation}
where $\rho_A=\tr_{BC}(\ketbra*{\widetilde{\psi}}{\alpha\beta\gamma})$.
The singular value decomposition of $\rho_A$, i.e., $\rho_A=UDV^\dagger$, provides the optimal choice $U_A=VU^\dagger$.
From this we obtain
\begin{equation}
  \max_{U_A}\abs{\tr_A(U_A\rho_A)}=\sum_i s_i(\rho_A).
\end{equation}
For given random initial choices for the states and local unitaries we iterate the optimizations until we reach a fix point.
Assuming we run the algorithm on sufficiently many random initial states we obtain the maximal overlap $\lambda$ with a high probability.
In low dimensional systems, i.e., two qubits per node and three nodes, the algorithm converges rather quickly resulting in the putative $\mu$ after order of ten iterations, regardless of the initial state. {Furthermore, we note that it sometimes happens that the algorithm only finds a local maximum for a fixed encoding. Hence an optimization over the encoding needs to be taken into account. Such a change of the encoding can be incorporated into the triangle by adding a local swap gate. For higher dimensional systems the overlap strongly depends on the encoding and also the number of possible encodings grows. In addition to that one also needs to take into account an optimization over possibly different dimensions of the sources.}

Finally, we note that upper bounds on the overlap of a pure state with pure states from the ITN can also be obtained analytically.
To illustrate the idea, consider first a bipartite system, where we would like to maximize the overlap of pure states $\ket{\psi}\in S$ in some subset with some target state $\ket{\tau}$.
If the target state has the Schmidt decomposition $\ket{\tau}=\sum_i t_i \ket{ii}$ and the Schmidt coefficients $s_i$ of states in the subset obey some  constraint $\{s_i\} \in \mathcal{S}$, then the overlap is bounded by
\begin{equation}
  \sup_{\ket{\psi}\in S} |\braket{\psi}{\tau}|^2 \leq \sup_{\{s_i\} \in \mathcal{S}} \Big| \sum_i s_i t_i \Big|^2
\end{equation}
For the case of the ITN with the distribution of qubits one can assume $\varrho_\alpha$, $\varrho_\beta$, and $\varrho_\gamma$ to be pure, having the Schmidt coefficients $[\cos(a), \sin(a)]$, $[\cos(b), \sin(b)]$, and $[\cos(c), \sin(c)]$, respectively.
Then, the above bound can be applied to all three bipartitions separately, and the best of these bounds can be taken.

To give a concrete example, let us consider the GHZ state $\ket{GHZ_2}=(\ket{000}+ \ket{111})/\sqrt{2}$, which has the same Schmidt coefficients for any bipartition.
Due to symmetry, we can assume for the Schmidt coefficients of the state $\ket{\psi}$ from the ITN that $\pi/4 \geq a\geq b\geq c \geq 0$.
Then, the overlap can be bounded by
\begin{equation}\label{eqn:maxf}
  \sup_{\ket{\psi}\in \mathrm{ITN}} |\braket{\psi}{GHZ_2}|^2\leq\frac12\max_{a,b,c}f(a,b,c)^2
\end{equation}
where
\begin{align}\label{eqn:f}
  f(a,b,c)=\min \big\{& \cos(a)\cos(b)+\sin(a)\cos(b), \\
    &\cos(c)\cos(b)+\sin(b)\cos(c), \nonumber \\
  &\cos(a)\cos(c)+\sin(a)\cos(c) \big\} \nonumber.
\end{align}
The minimization in Eq.~\eqref{eqn:f} corresponds to taking the optimal bound from all three bipartitions, and the maximization in Eq.~\eqref{eqn:maxf} is over all Schmidt coefficients for the ITN state.
After a short calculation, this gives the bound
\begin{equation}
  \sup_{\ket{\psi}\in \mathrm{ITN}} |\braket{\psi}{GHZ_2}|^2 \leq \cos(\frac{\pi}{8})^2 = \frac{2+\sqrt{2}}{4} \approx 0.8536.
\end{equation}
Similar calculations can be applied to other pure states.

\section{III. Intermediate scenario and Smolin state}
\label{app:smolin}

Here we explore shortly the scenario in which only the sources can be correlated.
This scenario is intermediate between the ITN and the CTN.
Here the states would take the form
\begin{equation}\label{eqn:triangle_cor}
  (U_A\otimes U_B\otimes U_C)\!\left(\sum_\lambda p_\lambda\varrho_\alpha^{(\lambda)}\!\otimes\varrho_\beta^{(\lambda)}\!\otimes\varrho_\gamma^{(\lambda)}\!\right)\!(U_A^\dagger\otimes U_B^\dagger\otimes U_C^\dagger).
\end{equation}
We conjecture that the set of states of this form is nonconvex.
While all such states belong to $\bigtriangleup_C$, we believe that the converse does not hold.
Moreover, one can show that there exist states of the form above that lie outside $\bigtriangleup_I$.

Consider a state $\rho_S$ on six qubits, or three ququarts equivalently, that corresponds to the unique $+1$ eigenstate of the two local operators $g_1=\sigma_x^{\otimes6}$ and $g_2=\sigma_z^{\otimes6}$, which are called the local stabilizers of $\rho_S$.
Alternatively we can also write
\begin{align}\label{eqn:smolinstate}
  \rho_S&=\frac{1}{32}\qty(\mathds{1}^{\otimes6}+g_1+g_2+g_1g_2)\\
  &=\frac{1}{32}\qty(\mathds{1}^{\otimes6}+\sigma_x^{\otimes6}-\sigma_y^{\otimes6}+\sigma_z^{\otimes6}).
\end{align}
This state corresponds, up to local unitary corrections, to the (generalized) Smolin state~\cite{Smolin2001,Augusiak2006}.
Here we show that this state is in $\bigtriangleup_C$ but not in $\bigtriangleup_I$.

Since the operators $\sigma_x^{\otimes2}$ and $\sigma_z^{\otimes2}$ commute, the result of Ref.~\cite[Lemma~1]{Wang2007} applies and shows that the state $\rho_S$ is separable with respect to any $2:2:2$ partition.
Hence it is separable in the sense that it could be produced by a single source sending Bell pairs to the parties $A$, $B$ and $C$.
Due to its permutation invariance --- the roles of all qubits are obviously interchangeable from the form of the stabilizers $g_1$ and $g_2$ --- the state is in $\bigtriangleup_C$.

Note that in this case, there is no need for correlations between the local unitaries as they can be chosen to be all trivial.
This shows that $\rho_S$ can be prepared in the intermediate scenario in which only the sources are correlated.
With the following proof that $\rho_S\notin\bigtriangleup_I$, we know that the separation between $\bigtriangleup_I$ and this intermediate scenario is strict.
Moreover, we conjecture that the set of state preparable in this configuration is nonconvex, which necessitates the separation with $\bigtriangleup_C$ to be also strict.

Going back to $\rho_S$, we now show that this state cannot be produced without the shared randomness between the sources.
This can be seen as an application of Observation~\ref{obs:rank}.
Suppose indeed that $\rho_S$ is of the form of Eq.~\eqref{eqn:triangle_ind}.
Since the global rank of the state is $16$ the only possible ranks for the states of the sources are either $1,4,4$ or permutations thereof, or $2,2,4$ and permutations thereof.
The first case always results in a state for which there exists a bipartition where the larger partition has rank $4$ which is not true for the Smolin state, which is maximally mixed on any bipartition.
In the second case one can always find a bipartition such that the larger partition has rank $8$ which is again not true for the Smolin state.
From that we conclude that it is not possible to prepare the Smolin state in the ITN.

\section{IV. Proof of Observation 4}

For the application of a single channel we start with a state $\varrho\in \bigtriangleup_I$ and remove the local unitaries. After applying two local channel, say on node $A$ and $B$, we can decompose the sum of the bipartite entropies as $S(AB)+S(BC)+S(AC)=S(A_\beta A_\gamma B_\gamma B_\alpha)+S(B_\gamma B_\alpha C_\alpha)+S(C_\beta)+S(A_\gamma A_\beta C_\beta)+S(C_\alpha)$ and the sum of the remaining entropies as given by $S(A)+S(B)+S(C)+S(ABC)=S(A_\beta A_\gamma)+S(B_\gamma B_\alpha)+S(C_\beta)+S(C_\alpha)+S(A_\beta A_\gamma B_\gamma B_\alpha C_\alpha C_\beta)$. Taking the difference results in $I_3(A:B:C)=I(A_\beta A_\gamma :B_\gamma B_\alpha)-I(A_\beta A_\gamma C_\beta:B_\gamma B_\alpha C_\alpha)\leq 0$ due to the the monotonicity of the mutual information, which itself is equivalent to the strong subadditivity condition of the von Neumann entropy. This concludes the proof.

\end{document}